\documentstyle[aaspp4,12pt,amssym,psfig]{article}
\begin{document}    
\newcommand{\be}{\begin{equation}}
\newcommand{\ba}{\begin{eqnarray}}
\newcommand{\ee}{\end{equation}}
\newcommand{\ea}{\end{eqnarray}}

\title{Lensing-Induced Structure of Submillimeter Sources:
Implications for the Microwave Background}
 
\author{    
Evan Scannapieco${}$, Joseph Silk${}$, and Jonathan C. Tan}
\affil{Departments of Physics and Astronomy,
University of California, Berkeley, CA 94720-7304}

\begin{abstract}
We consider the effect of lensing by galaxy clusters on the angular
distribution of submillimeter wavelength objects.  While lensing does
not change the total flux and number counts of submillimeter sources,
it can affect the number counts and fluxes of flux-limited samples.
Therefore imposing a flux cut on point sources not only reduces the
overall Poisson noise, but imprints the correlations between lensing
clusters on the unresolved flux distribution.  Using a simple model,
we quantify the lensing anisotropy induced in flux-limited samples and
compare this to Poisson noise.  We find that while the level of
induced anisotropies on the scale of the cluster angular correlation
length is comparable to Poisson noise for a slowly evolving cluster
model, it is negligible for more realistic models of cluster
evolution.  Thus the removal of point sources is not expected to
induce measurable structure in the microwave or far-infrared
backgrounds.
\end{abstract}

\keywords{gravitational lensing -- galaxies: clusters: general --
infrared: galaxies -- cosmic microwave background}

\newpage

\section{Introduction}

Due to the steep slope of the Rayleigh-Jeans portion of their
emission spectra, dusty galaxies have submillimeter
luminosities that are greatly enhanced as a function of
redshift.  This results in large negative submillimeter K-corrections,
causing the number of galaxies as a function of observed flux to be
roughly constant between redshifts 1 and 10 (Franceschini et al.\
1991; Blain \& Longair 1993).  Thus observations that can resolve
submillimeter sources at $z=1$ should be able to detect the full
history of these objects, and one would expect submillimeter
source counts to have a much steeper slope as a function of limiting
flux than source counts in other wavebands.

While this strong dependence on limiting flux had long been predicted
(Blain \& Longair 1993; Blain \& Longair 1996), the overall surface
density of these objects remained uncertain to within a factor of a
thousand until recently.
Using the the new Submillimeter Common User Bolometer
Array (SCUBA) (Holland et al.\ 1999) on the James Clerk Maxwell
Telescope, several groups (Barger et al.\ 1998; Eales et al.\ 1993;
Hughes et al.\ 1998; Smail, Ivison, \& Blain 1997) have 
detected objects at $850 \; \mu$m with fluxes of a few mJy.  Meanwhile,
parallel investigations utilising the FIRAS instrument on COBE (Puget
et al.\ 1996; Fixsen et al.\ 1998; Hauser et al.\ 1998), DIRBE
combined with IRAS (Schlegel, Finkbeiner and Davis 1998) and TeV gamma
ray constraints via pair production against the IR background (Biller
et al.\ 1998) have begun mapping out the extra galactic background
from $\sim 5$ to $1000 \; \mu$m wavelengths.
Thus the past few years have not only given us
our first look at the high-luminosity submillimeter universe,
but allowed us to place strong constraints on the unresolved distribution
as well.    This has allowed several authors
to investigate the effects of confusion noise
from unresolved sources both on galaxy surveys and measurements
of the cosmic microwave background (CMB) (Blain, Ivison, \& Smail 1998;
 Blain et al.\ 1998; Scott \& White 1998).

The steepness of submillimeter source counts also makes them highly
sensitive to the effects of gravitational lensing.  As
a gravitational lens amplifies the flux of objects behind it
while at the same time reducing the effective volume that is being
seen, only objects with sufficiently steep source counts
display a statistical increase in bright-end counts when lensed.  
Thus one would expect
the lensed fraction of bright submillimeter-wave sources 
to be much larger that the fraction in the optical and radio
wavebands (Blain 1996).  This sensitivity has motivated
several SCUBA searches for objects behind galaxy clusters, using
lensing to boost resolved source counts.  It
has also led to investigations of the effect of lensing by both
galaxies and galaxy clusters on overall number counts and the prospect
of measuring the degree of lensing from all-sky microwave experiments
(Blain 1996; Blain 1997; Blain 1998).

In addition to changing the number counts of bright objects, lensing
induces structure in the unresolved source distribution.  As the
clumping of matter cannot affect the total flux or number of sources,
lensing skews the flux distribution by making certain sources
brighter while at the same time preserving the total flux and number of
objects.  Thus imposing a flux cut causes a larger flux decrement in
the areas behind gravitational lenses.  As clusters of galaxies are
known to have large angular correlations (Dalton et al.\ 1984; Tadros,
Efstathiou, \& Dalton 1998), the removal of resolved point sources
will imprint a correlated decrement on the unresolved distribution of
submillimeter sources.

As future experiments such as the Planck satellite
(Bersanelli et al.\ 1996) will be making sensitive all-sky
measurements at microwave and submillimeter wavelengths and will use
flux-limited samples to investigate the structure of background
emission, it is important that this effect be quantified.  While
many elements contributing to this structure remain uncertain, current
cluster and submillimeter observations have improved our knowledge to
the point where this quantification is now possible.

In this work we develop a simple model of lensing by clusters,
appropriate to the degree at which cluster evolution is understood.
We then combine our lensing model with a semi-empirical model of the
submillimeter universe (Tan, Silk \& Balland 1999, hereafter TSB),
based on recent observations.  This leads to a quantification of
induced structure by lensing that is useful in determining the extent
to which this effect needs to be considered in future observations.

The structure of this work is as follows: In Section 2 we discuss
various models of the lensing of high redshift objects by clusters.
In Section 3 we examine the number and spectral evolution of the
submillimeter sources themselves and apply our lensing models to
predict the number counts, fluxes, and clustering properties of lensed
sources.  In Section 4 we compare these predictions 
to Poisson noise and discuss implications for background
experiments.  Conclusions are given in Section 5.

\section{Models of Gravitational Lensing}

\subsection{Lensing by Clusters}

The effect of gravitational lensing on the number densities 
of distant sources can be described by $F(A,z) dA$, 
the probability that a source at redshift $z$ will have
its flux density amplified by a factor between $A$ and $A + dA$.
For any distribution of objects 
\be
\int_0^\infty F(A,z) dA = 1 =
\int_0^\infty A F(A,z) d A,
\label{eq:conserve}
\ee
the first equality being a statement of conservation of
probability, and the second following from flux conservation.
Note that while naively one would expect lensing to cause only amplification,
the requirement that the mean flux density be unaffected by the clumping
of matter in the universe
results in a net deamplification of unlensed objects 
(Weinberg 1976; Dyer \& Roeder 1973).
Thus for all redshifts $F(A,z)$ is a skewed distribution
with a maximum slightly less than one and a long tail corresponding 
to strong lensing events.

By integrating over the comoving mass distribution of clusters, $n(M,z)$, 
$F(A,z)$ can be determined as
\be
F(A,z) = \pi \frac{d}{d A} \int_0^z dz' \int_0^\infty dM b(A,M)^2 
              (1 + z')^2 \frac{dn(M,z')}{d M} \frac{dr}{dz'}, 
\label{eq:F}  
\ee
where $dr$ is a comoving radial distance element,
and $b$ is the proper impact parameter for a source at redshift
$z$ and a lens mass $M$ located at a redshift $z'$ (Peacock 1982).

The proper impact parameter is the angular size distance to the
lens times the angle of deflection by that lens, and 
is dependent on the geometry of the lensing distribution.
Fortunately, the most simple assumption, the $\rho \propto r^{-2}$
isothermal sphere distribution, has been shown both theoretically and
observationally to be a good fit for intermediate radii between 0.1
and 1.5 $h^{-1}$ Mpc, where $h$ is the Hubble constant in units
of 100 ${\rm km}\,{\rm s}^{-1}\,{\rm Mpc}^{-1}$
(Tyson \&
Fischer 1995; Navarro, Frenck \& White 1996).  For large magnifications
this model gives (Peacock 1986)
\be
b(A,M) = 
\frac{8 \pi \sigma^2_v}{c^2} 
\frac{D(0,z') D(z',z)}{D(0,z)} \frac{1}{A},
\label{eq:b}
\ee
were $\sigma_v$ is the velocity dispersion of the lens,
$D(0,z')$ is the angular size distance to the lens,
$D(0,z)$ is the angular size distance to the lensed object,
and $D(z',z)$ is the angular size distance from the lensed
object to the lens.  For clusters, 
$\sigma_v = 860 \, m^{1/2}$km/sec, where $m$ is the mass of the cluster
within $1.5 \, {\rm Mpc} \,h^{-1}$ in units of 
$6 \times 10^{14} \, M_\odot \, h^{-1}$.
Strictly speaking, the angular size distances in Eq.\ \ref{eq:b}
are the Dyer-Roeder distances which account for the inhomogeneity
of the lensing objects.  For our purposes here, however, we
use the homogeneous distances (Young et al.\ 1980). 

Bahcall, Fan, \& Cen (1997) have used the CNOC/EMSS sample of
high-redshift clusters to determine the observed evolution of cluster
abundances.  For massive clusters with $m \ge 1.05$, the evolution
in redshift is given by $\log n(z) \propto
-z$ with $\log (n(z=0)/n(z=0.5)) = 0.7 \pm 0.3(1 \sigma)$
(Fan, Bahcall \& Cen 1997, hereafter FBC).  Carlberg
et al.\ (1996) give similar results.  One could envision modelling
the evolution of clusters from a Press-Schechter (Press \& Schechter
1974) viewpoint, counting the number of peaks above a certain
threshold as a function of redshift and cosmology.  The problem with
this approach, however, is that while the evolution of the number
density is only weakly dependent on the cosmological parameters, it is
strongly dependent on the unknown amplitude of mass fluctuations on
the 8 $h^{-1}$ Mpc scale, $\sigma_8$ (FBC).  As
$\sigma_8$ is uncertain from observations, a simple fit
to data is the best we can hope for even if we fix the other
cosmological parameters.

\begin{figure}
\centerline{ 
\psfig{file=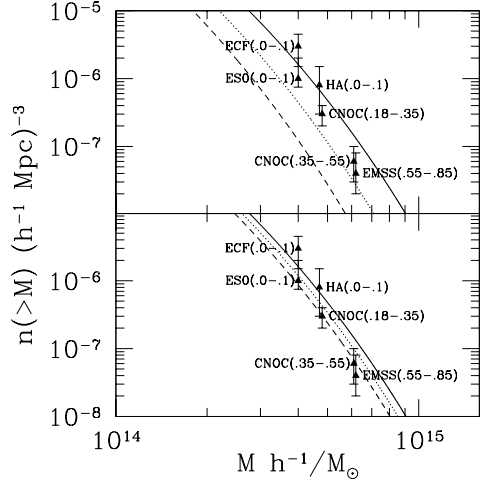,width=2.6in}}
\caption{The $n(>M)$ relation as calculated fit to the data compiled
by Carlberg et al.\ (1997), with redshift ranges indicated
and acronyms are as described in Sec 2.1.  
The solid lines are $z=0$,
 the dotted lines are $z = 0.5$, and the dashed lines are $z = 1.0$.
The upper panel shows the FBC ($\alpha = 2.9$) model and the
lower panel shows the slowly-evolving ($\alpha = 0.7$) model.
$M$ is the mass of the cluster within $1.5 h^{-1}$ Mpc.
}
\label{fig:clustermodel}
\end{figure}

For our purposes then, we adopt a simple model for the number
and redshift evolution of clusters that retains the Press-Schechter
form but is fit to the observed number density and evolution of clusters
\be
\frac{dn}{dm} = n_0 m^{-7/4} e^{- (m^{1/2}(z\alpha+\beta))}.
\ee
Here the $m^{1/2}$ dependence in the exponential and overall
$m^{-7/4}$ dependence corresponds to a mass density power spectrum
$P(k) \propto k^{-1.5}$ around 8 $h^{-1}$ Mpc, where cold dark matter
models give $P(k) \propto k^{-1.4}$ (Bardeen et al.\ 1986) We fix
these parameters to the $n(>M)$ relation for clusters as calculated in
Carlberg et al.\ (1997), which used the Canadian Network for
Observational Cosmology (CNOC) (Yee, Ellingson, \& Carlberg 1996)
cluster sample drawn from the Extended Medium Sensitivity Survey
(EMSS) (Henry et al.\ 1992) as well as the ESO Cluster Survey (Mazure
et al.\ 1996) and other cluster data sets (Henry \& Arnaud 1991,
hereafter HA; Eke, Cole, \& Frenck 1996, hereafter ECF).  This
comparison yields $\beta = 10$ and $n_0 \times e^{-\beta} = 10^{-6}
h^{-3} {\rm Mpc}^{-3}$.  For this set of parameters the $\log
(n(z=0)/n(z=0.5))$ comparison of FBC quoted above corresponds to
$\alpha = 2.9 \pm 1.1 (1 \sigma)$.  Note, however, that cluster
evolution fits have been carried out by other authors and there is
some disagreement between results (Blanchard \& Bartlett 1998; Sadat,
Blanchard, \& Oukbir 1998).  In Fig.\ \ref{fig:clustermodel} we
compare the data compiled in Carlberg et al.\ (1997).  with the
``best-fit'' FBC, $\alpha = 2.9$ evolution model and a slowly
evolving, $\alpha = 0.7$ model that we consider in further detail
below.

Modelling clusters in this
manner, we find that for large amplifications
$F(z,A) = a(z) A^{-3}$ where
\be
a(z) = 0.007
\int_0^z  dz'  
\frac{dr}{dz'} \frac{S(\eta(z'))^2 S(\eta(z-z'))^2}{S(\eta(z))^2} 
\int_{m_{\rm min}}^{m_{\rm max}} dm \, m^{1/4} \, e^{10} 
\, e^{- (m^{1/2}(z\alpha + 10))},
\ee
where $\eta$ is a dimensionless comoving distance ($\eta = r H/c$), 
$S(\eta) = \eta$ in a flat universe, and
$S(\eta) = \sinh( \eta \sqrt{1 - \Omega_{\rm M} - \Omega_\Lambda})
/\sqrt{1 - \Omega_{\rm M} - \Omega_\Lambda}$ in an open universe.
Note that this expression is comparable to the model considered by (Pei 1995), 
which he derived from observations of the empirical distribution of dense
groups and rich clusters (Zabludoff et al.\ 1993), although our approach
is quite different.  Throughout this paper we fix 
$M_{\rm min} = 3 \times 10^{14} M_\odot$ and 
$M_{\rm max} = 1 \times 10^{15} M_\odot.$

\begin{figure}
\centerline{ 
\psfig{file=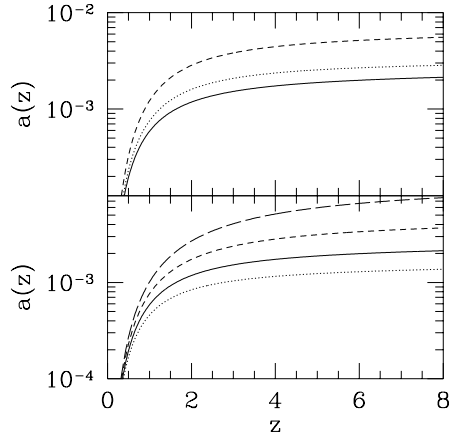,width=2.4in}}
\caption{
Upper panel: Plots of $a(z)$, a measure of the probability of strong
lensing as function of redshift, for three different cosmological
models: a flat, $\Omega_{\rm M} = 1$ universe (solid), an open model
with $\Omega_{\rm M} = 0.3$ (dotted), and a flat cosmological constant
model with $\Omega_{\rm M} = 0.3$ (dashed). In all cases $\alpha$ =
2.9.  
Lower panel: Plots of $a(z)$ for Einstein-de Sitter cosmologies with
four different models for the evolution of clusters: $\alpha = 4.0$
(dotted), $\alpha = 2.9$ (solid), $\alpha = 1.8$ (short-dashed) and
$\alpha = 0.7$ (long-dashed).}
\label{fig:aofz}
\end{figure}

In Fig.\ \ref{fig:aofz} we plot this function for three different
cosmologies in the upper panel, and for four different cluster
evolution models in the lower panel.  Notice that at moderate
redshifts the cosmological constant model gives slightly more than
twice the lensing of the Einstein-de Sitter model, while the
slow-evolution model of lensing is four times higher than the FBC
evolutionary model.  Thus while the degree of lensing is quite
sensitive to cosmological parameters (Carroll \& Press 1992), lensing
by galaxy clusters is even more sensitive to the evolution of the
cluster population itself.

\subsection{Clustering of Clusters}

The effect of gravitational lensing on the clustering of a
distribution of sources depends on the probability that two sources at
redshift $z_1$ and $z_2$ separated by an angle $\theta$ will be lensed
by factors between $A_1$ and $A_1+ dA_1$ and $A_2$ and $A_2+ dA_2$:
$F(A_1,z_1,A_2,z_2,\theta) dA_1 dA_2$.  Similar probability and flux
conservation relations to Eq.\ \ref{eq:conserve} hold for this
distribution:
\be
\int_0^\infty dA_1
\int_0^\infty dA_2
F(A_1,z_1,A_2,z_2,\theta) = 1 =
\int_0^\infty A_1 dA_1
\int_0^\infty A_2 dA_2
F(A_1,z_1,A_2,z_2,\theta).
\ee
In this case
$F(A_1,z_1,A_2,z_2,\theta)$ can be determined as
\ba
F(A_1,z_1,A_2,z_2,\theta) &=& 
\frac{d}{d A_1} \frac{d}{d A_2}
 \int_0^{z_1} \frac{dr}{d z_1'}
dz_1' (1+z_1')^2  \times \nonumber \\
& & \qquad \qquad \int_0^{z_2} \frac{dr}{d z_2'}
dz_2' (1+z_1')^2 
\pi b(A_1)^2 \pi b(A_2)^2
n^2(z_1,z_2,\theta),
\ea
where $n^2(z_1,z_2,\theta)$ is the product of the
 number density of clusters at redshifts $z_1$ and $z_2$
separated by an angle $\theta$. 

Strictly speaking 
\be
n^2(z_1,z_2,\theta) =
n(z_1)n(z_2)[1+\xi(r(z_1,z_2,\theta))],
\ee
where $\xi(r)$ is the cluster spatial correlation function,
 and $r(z_1,z_2,\theta)$ is the comoving distance
between two clusters as a function of their redshifts and
separation.
There are however a number of difficulties in implementing such an
approach, as not only is $r$ dependent on unknown cosmological
parameters, but the spatial correlation function of clusters is
subject to a number of observational uncertainties.  While several
authors have fit the optical cluster-cluster correlations with a power
law of the form $\eta = (r/r_0)^{-\gamma}$ with $-2.2 \leq \gamma \leq
-1.8$ (Peacock \& West 1992; Dalton et al.\ 1994), the
cluster-cluster correlation length $r_0$ has remained controversial.
Here values are much more uncertain,
ranging from 15 $h^{-1}$ Mpc (Nichol et al.\ 1992; Dalton et
al.\ 1994) to 25 $h^{-1}$ Mpc (Peacock \& West 1992; Abadi, Lambas, \&
Muriel 1998), with various dependences on cluster mass and mean
intercluster separation (Bahcall \& Soneira 1983; Bahcall \& Cen
1992).  To compound this problem, the clustering of clusters should be more
severe at higher redshifts due to the fact that the first density
peaks to collapse were more rare than those collapsing recently.

Due to these uncertainties, we avoid implementing the spatial
correlation function and approximate the number density of 
objects as
\be
n^2(z_1,z_2,\theta) \approx 
n(z_1)n(z_2)[1+w(\theta)],
\ee
with $w(\theta) = (\theta/\theta_0)^{1+\gamma}$, with 
$\theta_0 = 1.1^\circ$ observed in the APM survey
(Dalton et al.\ 1994).  Note however, smaller values of $\theta_0$ 
may be more appropriate for lensing due to the larger redshift
range at which clusters have an observable effect.
Thus our choice for the angular correlation function of
lensing clusters should be considered an upper limit.

As the form of the peak of $F(A,z)$ at amplifications 
slightly less than unity has little effect on the results
of lensing, Blain (1996), following Vietri \& Ostriker (1982), examined
a simple model for $F$:
\be
F = H(z) \delta(A - A_0(z)) + A^{-3} a(z) \theta(A-A_{\rm min})
\theta(A_{\rm max} - A),
\label{eq:Fdelta}
\ee 
where $H(z)$ and $A(z)$ are fixed by the conservation relations
Eq.\ \ref{eq:conserve}, and
$\theta$ is the Heaviside step function,
which is used to exclude low magnifications at which
we would expect Eq.\ \ref{eq:F} to break down and
high magnifications at which the isothermal sphere approximation
is invalid.
Working to first order in $a(z)$, this gives
\be
H(z) =
1 - a(z)
\left[
{\frac{A_{\rm min}^{-2} - A_{\rm max}^{-2}}{2}} \right],\qquad \qquad
A_0(z) =  1 - a(z) \left[\frac{2 A_{\rm min} - 1}{2 A_{\rm min}^2}
                       -\frac{2 A_{\rm min} - 1}{2 A_{\rm max}^2} \right].
\label{eq:handa}
\ee
Physically, this corresponds to
modelling each cluster as an
isothermal sphere in which we have truncated the lens
profile at the angular size distance corresponding to an
amplitude of $A_{\rm min}$, smoothed the distribution of
matter within a core radius corresponding to $A_{\rm max}$, 
and imposed a net demagnification
of objects relative to their amplitudes in a homogeneous universe.

In order to study the clustering properties of lensed objects,
we examine a similar model for the two point amplification function
\ba
F(A_1,z_1,A_2,z_2,\theta)  
&=&  H(z_1,z_2,\theta) 
\delta(A_1- A_0(z_1,z_2,\theta)) 
\delta(A_2- A_0(z_1,z_2,\theta)) 
\nonumber \\
&+& A_1^{-3} A_2^{-3} 
\theta(A_1 - A_{\rm min})
\theta(A_2 - A_{\rm min}) \times \nonumber \\
& & 
\theta(A_{\rm max} - A_1)
\theta(A_{\rm max} - A_2)
a(z_1)a(z_2)[1+ w(\theta)],
\label{eq:Ftheta}
\ea
where $H$ and $A_0$ are fixed by conservation of probability 
at all values of redshift and angular separation. Imposing these conditions
gives us
\ba
H(z1,z2,\theta) &=& 
1 - a(z_1)a(z_2) [1+w(\theta)]
\left[
\frac{1}{2 A_{\rm min}^2} -
\frac{1}{2 A_{\rm max}^2}
\right]^2,
\nonumber \\
A_0(z1,z2,\theta) &=& 
1 - 
a(z_1)a(z_2) \frac{1+w(\theta)}{2}
\left[
\left(
\frac{1}{A_{\rm min}} -
\frac{1}{A_{\rm max}} \right)^2
-
\left(
\frac{1}{2 A_{\rm min}^2} -
\frac{1}{2 A_{\rm max}^2} \right)^2
\right].
\label{eq:handatheta}
\ea

Fixing the typical unlensed size of a submillimeter source to be 
$25 h^{-1} {\rm kpc}$
and the redshifts of a typical cluster and submillimeter source
to be $0.5$ and $3$ respectively, gives a value of $A_{\rm max} = 70$
in a flat universe ($\Omega_{\rm M} = 1$), with higher values for open
cases. Throughout this work we fix $A_{\rm max} = 50$, as a 
conservative limit.
Somewhat more arbitrarily, we have set
a minimum threshold of $A_{\rm min} = 2$, because it is a round number
above which more sophisticated lensing models typically display 
$F(A,z) \propto A^{-3}$ profiles (Peacock 1982; Pei 1995).
We have examined the effect of these cutoffs on our model of cluster
lensing and find that changing them introduces less than 20\% uncertainty
in our results for reasonable value of $A_{\rm max}$ and $A_{\rm min}$
and relevant threshold fluxes.

\section{Submillimeter Sources}

\subsection{A Semi-Empirical Model}

We use the simple model of TSB for the evolution of sources in the
infra-red and submillimeter.  The model evolves the properties of the
local infra-red and submillimeter sources backwards in time,
accounting for dust, gas, and spectral evolution.  It agrees well with
observations and constraints of the extra-galactic background from
$\sim 5$ to $1000\; \mu$m (Puget et al.\ 1996; Fixsen et al.\ 1998;
Hauser et al.\ 1998; Biller et al.\ 1998), as well as with number
count surveys from ISO (Puget et al.\ 1999; Aussel et al.\ 1998;
Altieri et al.\ 1998) and SCUBA (Smail et al.\ 1998).  Two types of
sources, disk galaxies and starbursts, have significant fluxes in the
submillimeter regime due to the reprocessing of stellar radiation by
dust.

Disk submillimeter luminosity evolution is derived from observations
of the star formation rate and metallicity histories of the Milky Way.
A Schmidt Law relation (Kennicutt 1998) between star formation rate
and gas density is used to derive an optical depth history.  The
contributions from both young and old stars to dust heating are
modelled.  The number evolution of disks is obtained from models of
collision-induced galaxy formation (Balland, Silk, \& Schaeffer 1998)
which attributes the formation of different morphological types to
galaxy-galaxy interactions.

We model starburst evolution assuming they result from the mergers of
gas-rich disks. The typical luminosity increases with the gas
fraction, but decreases with the mean mass of the galaxies.  The
starburst number density is proportional to the instantaneous
galaxy-galaxy collision rate.

These evolutionary models are applied to the present day far infra-red
luminosity function, as observed by IRAS from 40 to 120 $\mu$m
(Saunders et al.\ 1990), assuming an Einstein-de Sitter cosmology with
$h = 0.5$.  Given the uncertainties in the lensing model due to the
evolution of clusters, we have not extended the submillimeter model to
other cosmologies, but rather we restrict our attention to this
cosmology.  As lensing is more severe in open models and models with
nonzero $\Omega_\Lambda$, these are expected to yield results above
the $\alpha = 2.9$, $\Omega_{\rm M} = 1$ model, although we still
consider the $\alpha = 0.7$, $\Omega_{\rm M}=1$ model we examine below
to be an overestimate of lensing even compared to 
$\Omega_{\rm M} < 1$ cosmologies.

For simplicity, a cut is made in 
luminosity at $L_{\rm cut}\sim 6\times 10^{10}\:h^{-2}\:L_{\odot}$, 
below which sources are treated as disks, and above which they are treated 
as starbursts (Sanders \& Mirabel 1996).
The results show the most sensitivity to this parameter, and to illustrate 
model uncertainties, we investigate the effect of varying $L_{\rm cut}$
in Sections 3.2 and 4.

The luminosity function is also divided empirically into 
different spectral classes (Malkan \& Stecker 1998), and these 
spectra are analytically 
extended beyond 400 $\mu$m into the submillimeter regime, 
assuming a dust emissivity dependence of $\nu^{\beta}$.
Here we fix $\beta=1.5$ (Franceschini, Andreani, \& Danese 1998; 
Roche \& Chandler 1993), but our results are quite insensitive to
this value as most of the sources detected are at $z \gtrsim 1$.

\subsection{Lensed Submillimeter Sources}

Applying  Eq.\ \ref{eq:Fdelta} and \ref{eq:handa} to our model
for number counts of submillimeter objects as a function of
flux and redshift, we are able to calculate the perturbation
to the number of sources per unit flux due to lensing by clusters.
If we take $\frac{dN_{\rm total}}{d S} = \frac{dN_{\rm unlensed}}{d S} +  
\Delta \frac{dN}{d S} $ then 
\ba
\Delta \frac{d N}{d S} = & &
\left[         \frac{2 A_{\rm min} -2}{2 A^2_{\rm min}} 
              -\frac{2 A_{\rm max} -2}{2 A^2_{\rm max}} 
\right]
\frac{d N_L}{d S} \nonumber \\
        & + &
\left[
 \frac{2 A_{\rm min} - 1}{2 A^2_{\rm min}}  
-\frac{2 A_{\rm max} - 1}{2 A^2_{\rm max}}  
\right]
S \frac{d^2 N_L}{d^2 S} 
+\int_{A_{\rm min}}^{A_{\rm max}} 
 \frac{dA}{A^4}  \frac{dN_L}{dS} (S/A),
\label{eq:dndslens}
\ea
where 
$\frac{d N_L}{d S} \equiv \int_0^\infty a(z) \frac{dN}{dS dz}{dz}$.
Notice that both 
$ \int^\infty_0 dS \Delta \frac{d N}{dS }$
and $\int^\infty_0 SdS \Delta \frac{d N}{d S}$ are zero,
as the lensing of objects does not change the total number of observed
objects or the total integrated flux from those objects.  
Suppose, however, we consider only objects above a certain threshold
flux $S_{\rm cut}$. $\int_{S_{\rm cut}}^\infty dS \Delta \frac{d N}{d S}$
and $\int_{S_{\rm cut}}^\infty S dS \Delta \frac{d N}{d S}$ need
not be zero, as both the total number of objects above a given
threshold and the total
contribution to the flux  by these objects can vary according to the
degree of lensing.  Furthermore, the total change above
a threshold must be equal to the change below that threshold.  
Thus
\be
\int_{S_{\rm cut}}^\infty dS \Delta \frac{d N}{d S} = 
-\int^{S_{\rm cut}}_0 dS \Delta \frac{d N}{d S},
\qquad
{\rm and}
\qquad
\int_{S_{\rm cut}}^\infty S dS \Delta \frac{d N}{d S} = 
-\int^{S_{\rm cut}}_0 S dS \Delta \frac{d N}{d S},
\label{eq:sfluxcon}
\ee
for all values of $S_{\rm cut}$, and we are free to consider these
quantities interchangeably.  

By applying Eq.\ \ref{eq:Ftheta} and \ref{eq:handatheta}, the
perturbation to $\frac{d^2 N^2}{d S_1 d S_2}$ due to lensing 
can be similarly calculated.  Here we find 
\ba
\Delta \frac{d^2 N^2}{d S_1 d S_2} & =&
(1 + w(\theta))
\left\{
\left[
\left(
\frac{1}{A_{\rm min}} -
\frac{1}{A_{\rm max}} \right)^2
- 2
\left(
\frac{1}{2 A_{\rm min}^2} -
\frac{1}{2 A_{\rm max}^2} \right)^2
\right] 
 \frac{d N_L}{d S_1} \frac{d N_L}{d S_2} \right. \nonumber \\
& &+ \frac{1}{2}
\left[
\left(
\frac{1}{A_{\rm min}} -
\frac{1}{A_{\rm max}} \right)^2
- 
\left(
\frac{1}{2 A_{\rm min}^2} -
\frac{1}{2 A_{\rm max}^2} \right)^2
\right] 
        \left( S_1 \frac{d^2 N_L}{d S_1^2}  \frac{d N_L}{d S_2} 
        +  \frac{d N_L}{d S_1} S_2 \frac{d^2 N_L}{d S_2^2} \right) 
  \nonumber \\
& & + \left.
\left( 
\int_{A_{\rm min}}^{A_{\rm max}} 
\frac{dA_1}{A_1^4}  \frac{dN_L}{dS_1} (S_1/A)
\right)
\left(
\int_{A_{\rm min}}^{A_{\rm max}} 
\frac{dA_2}{A_2^4}  \frac{dN_L}{dS_2} (S_2/A)
\right)
\right\}.
\label{eq:dndssquare}
\ea

We now restrict our attention to flat cosmologies with $h = 0.5$.
Given the uncertainties in our knowledge of clusters and submillimeter
sources, we expect this model to be sufficiently representative to
examine the importance of structure induced by lensing.  Due to the
large volume and degree of lensing in open and cosmological constant
models, we expect $\alpha = 2.9$, $\Omega_{\rm M} < 1$ models to
fall somewhere between our Einstein-de Sitter $\alpha = 2.9$ and
slowly evolving cluster models.

\begin{figure}
\centerline{ 
\psfig{file=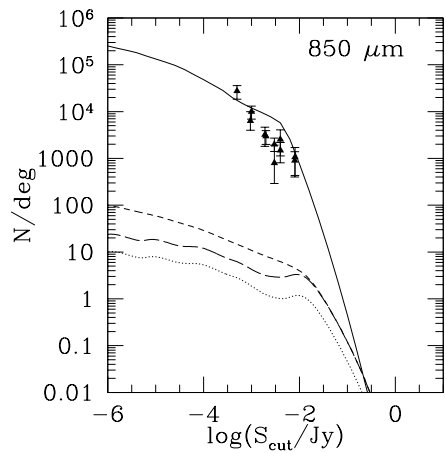,width=2.4in}}
\caption{
Number density of resolved objects as function of flux threshold at
850 $\mu$m.  The solid line is the total number density of objects,
$N(>S_{\rm cut})$, from TSB. Note that his particular model is not
fine tuned to match the sub-mm observations, but rather is one example
of a range of possible models.  The lower curves show the perturbation
due to lensing.  For the $\Omega_{\rm M} = 1$, $\alpha = 0.7$ model
the short-dashed line shows the change in number densities due to
strong lensing by clusters, $\Delta N(>S_{\rm cut})$, and the
long-dashed line is the number of strongly lensed objects (the
contribution due to the last term of Eq.\ \protect\ref{eq:dndslens}).
The dotted curve shows $\Delta N(>S_{\rm cut})$ for the FBC
Einstein-de Sitter model with $\alpha = 2.9$.  The points are the
SCUBA data as tabulated by Smail et al.\ (1998).  All errorbars are $1
\sigma$.  }
\label{fig:sc}
\end{figure}

In Fig.\ \ref{fig:sc} we plot the total number of objects above a
given threshold flux as a function of $S_{\rm cut}$ as well as the
change due to lensing by clusters at 850 $\mu$m, roughly corresponding
to the SCUBA observations and one of the high-frequency channels on
the Planck satellite.  The model shown in this figure is not fine tuned to
match the sub-mm observations, but rather is one example of a range of
possible models.  While the number of strongly lensed
objects is equal to $\Delta N(>S_{\rm cut})$ for large values $S_{\rm cut}$,
deamplification causes the net perturbation to $N(>S_{\rm
cut})$ to be an order of magnitude smaller than the number of strongly
lensed objects at lower $S_{\rm cut}$ values.

As noted by Peacock (1982), gravitational lensing is most important
when $\frac{d N}{d S}$ is steeper than $S^{-3}$ 
and is able to overcome the $A^{-3}$ dependence
of the high-magnification tail of the lensing distribution.
For more shallow distributions, however, the degree of lensing is smaller
by several orders of magnitude, and can even result in a net
decrement to $N(>S_{\rm cut})$ for sufficiently shallow slopes
(Broadhurst, Taylor, \& Peacock 1995).
Thus the difference between the number of strongly lensed objects and
$\Delta N(>S_{\rm cut})$ becomes important at values roughly 
corresponding to the turnover in the number counts of submillimeter objects.
In the $\alpha = 2.9$ and slowly evolving models $\Delta N(>S_{\rm cut})$
is almost of the same form, 
but is smaller for the slowly evolving model by a factor $\sim 3$.

\begin{figure}
\centerline{ 
\psfig{file=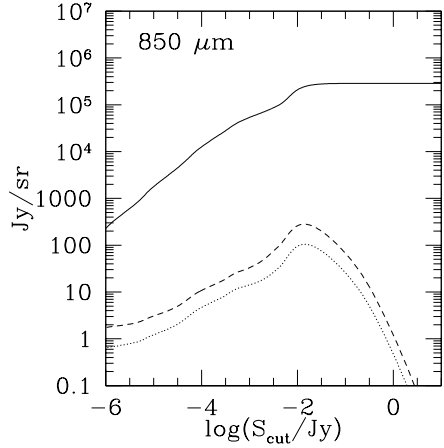,width=2.4in}}
\caption{
Flux per unit steradian in 
unresolved objects as a function of flux threshold at 850 $\mu$m.  The solid
line is unlensed flux, the dashed line is the 
decrement in flux due to lensing by clusters in the slowly evolving model
($\Omega_{\rm M} = 1, \alpha = 0.7$) and the dotted line is the flux decrement
in the $\Omega_{\rm M} = 1, \alpha = 2.9$, model.
}
\label{fig:flux}
\end{figure}

In Fig.\ \ref{fig:flux} we compare flux per unit angle due to
unresolved objects in lensed and unlensed distributions as a function
of flux cut at 850 $\mu$m.  As objects behind gravitational lenses are
fewer and more luminous, imposing an overall flux cut causes a
stronger decrement in the lensed distribution of objects than in the
unlensed case.  Again, this flux decrement in a slowly evolving model
is of the same form as in the FBC model, but scaled down by a factor
$\sim 3$.  Note that while the flux of lensed objects is only $\sim 0.1 \%$
of the total flux, the correlations induced by clustering 
can make this contribution important at large angular scales.

This discrepancy in fluxes also raises the question of the clustering
of the objects themselves.  While this is likely to be weak and at
very small angles, Scott \& White (1998) have demonstrated that this
clustering may be dominant over Poisson noise at angles $\sim 2'$,
although their model is only for illustrative purposes.  As this
contribution goes down as a function of flux-cut, and is important
only at small angles, for our purposes it is best thought of as an
uncertainty in the small-scale noise and will turn out not to affect
our results.

\begin{figure}
\centerline{ 
\psfig{file=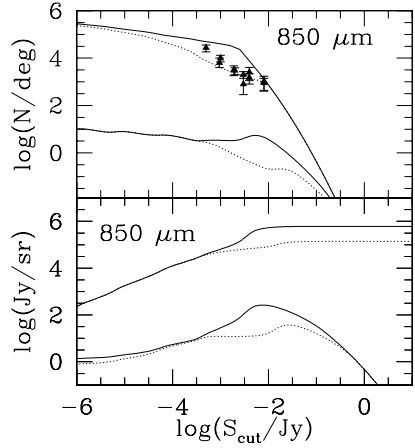,width=2.4in}}
\caption{
The effect of uncertainties in the submillimeter model
on lensing.  In the upper panel,
the upper curves show the total number of objects
without lensing by clusters, 
$N(>S_{\rm cut})$, for 
two different models while the lower curves
show the change in this number density due to lensing, 
$\Delta N(>S_{\rm cut})$. Here
$L_{\rm cut} = 3 \times 10^{10} h^{-2} L_\odot$ 
is represented by the solid lines, and the dotted lines
 represent $L_{\rm cut} = 12 \times 10^{10} h^{-2} L_\odot$. 
In the lower panel, the upper curves show total unlensed fluxes
and the lower curves show the flux decrement due to lensing,
with the lines corresponding to models as in the upper panel.
}
\label{fig:models}
\end{figure}

Our models are quite sensitive to the threshold between starburst and
disk galaxies.  Raising the value of $L_{\rm cut}$ suppresses the
high-$z$ luminous sources, shifting the turnover point in the
$\frac{dN}{dS}$ curve.  As the lensed flux is extremely sensitive to
this slope, the number counts and flux decrements in these model
differ dramatically.  In Fig.\ \ref{fig:models} we consider two
submillimeter models that overpredict and underpredict this slope.
The steep ($L_{\rm cut} = 3 \times 10^{10} h^{-2} L_\odot$) model
displays a large peak in the number counts and flux decrements that is
absent in the more shallow ($L_{\rm cut} = 12 \times 10^{10} h^{-2}
L_\odot$) model.  While these are extreme models, chosen to make a
point, ignorance of this slope is nevertheless one of the largest
uncertainties in our analysis.

\section{Background Anisotropies}

Contributions of point sources to background anisotropies can 
arise in two different ways: Poisson noise due to the intrinsic 
graininess of the distribution, and structured noise due to
the placement of the sources on the sky.
The first source of noise is given by
\be
C_\ell^I 
= \int_0^{S_{\rm cut}} dS S^2\frac{d N_{\rm Tot}}{d S} =
\int_0^{S_{\rm cut}}  dS S^2 
\left( \frac{d N}{d S} +
\Delta \frac{d N}{d S}
\right),
\ee
where we use the notation $C_\ell^I$ to denote the angular power 
spectrum in units of intensity rather than the usual decomposition
in terms of the temperature fluctuations. $C_\ell^I$ is related
to the usual $C_\ell$ by
\be
C_\ell = C_\ell^I \left( T_{\rm CMB} \frac{\partial B_\nu}{\partial T}
\right)^{-2} = C_\ell^I
\left(
2.71 \times 10^{10} \; {\rm mJy \, sr}^{-1} 
\frac{x^4 e^x}{(e^x -1)^2}
\right)^{-2},
\ee
where $x \equiv h \nu/ k_B T_{\rm CMB} = \nu/56.84$ GHz and $1 \; {\rm
Jy} = 10^{-26} \;{\rm W} {\rm m}^{-2} {\rm Hz}^{-1}$ (Scott \& White
1998).  Note that here we follow the conventional notation for
microwave background fluctuations, expressing our results in terms of
$C_\ell$s, given by the Legendre transform of the correlation function
$C(\theta)$.

\begin{figure}
\centerline{ 
\psfig{file=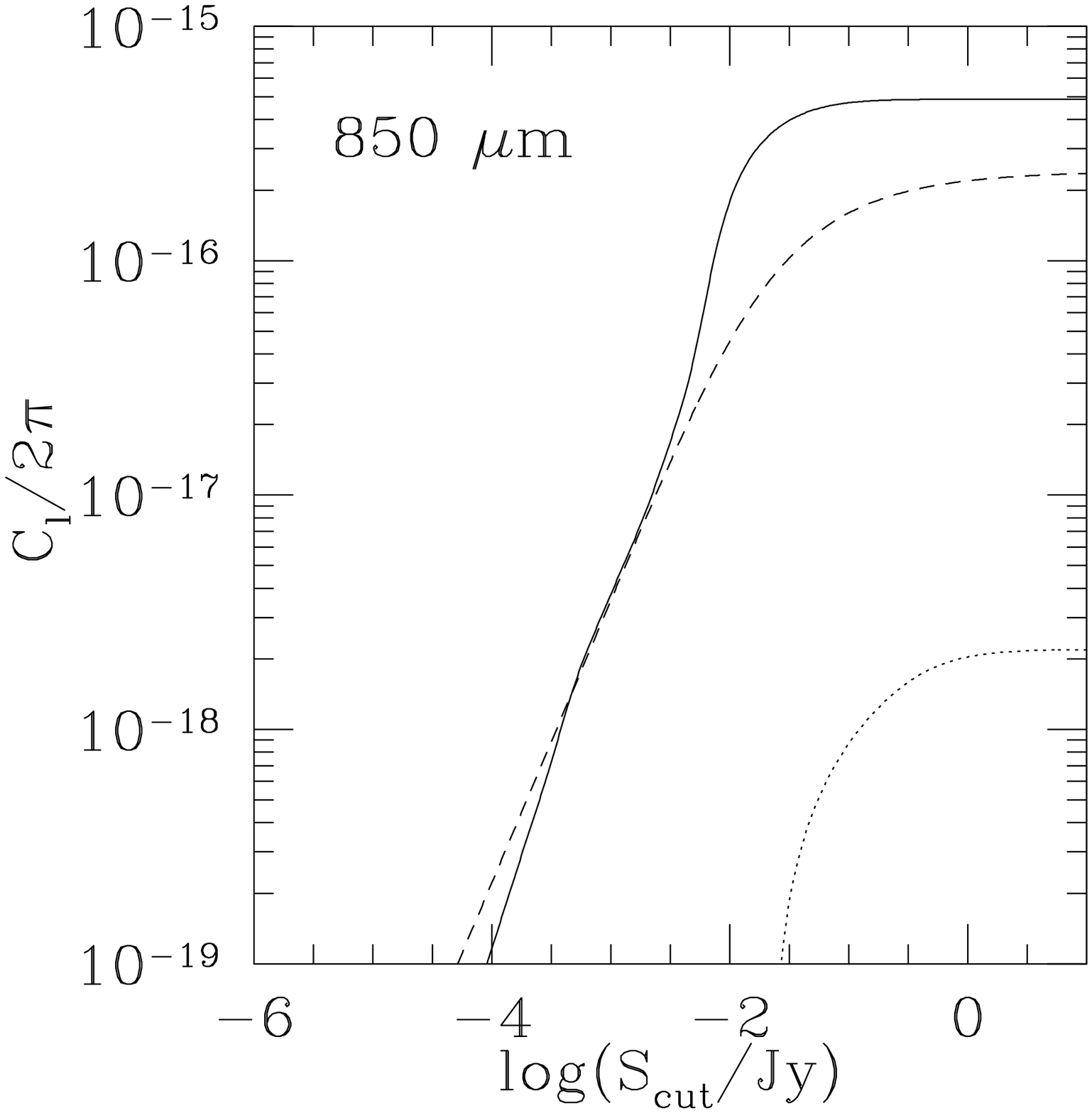,width=2.4in}}
\caption{
Contribution of confusion noise to measurements of
$C_\ell/2\pi$ as a function of flux cut
The solid
line is unlensed flux while the dotted line is the 
increment in this flux due to lensing by clusters.
The dashed line is the Poisson noise as calculated from a simple
double-power law fit by Scott \& White (1998).
}
\label{fig:poisson}
\end{figure}

As lensing skews the distribution by gathering the flux into
fewer, brighter objects, it does slightly increase the Poisson noise
of the observations.  This contribution to the anisotropies is a 
single-point process, however, with no intrinsic scale, and is completely 
independent of the placement of the lensing objects on the sky.
The level of Poisson noise in our model and the increment due
to lensing are shown in Fig.\ \ref{fig:poisson}.
Notice that removing point sources above a certain flux only decreases
the Poisson noise, as it reduces the total power in the grainy distribution.

The second contribution of point sources to anisotropies, that of
inhomogeneities in the mean flux as a function of position is given 
by
\ba
C(\theta)^I = 
 \int_0^{S_{\rm cut}} S_1 d S_1  
 \int_0^{S_{\rm cut}} S_2 d S_2
 \left[
 \Delta \frac{dN^2 (S_1,S_2,\theta)}{d S_1 d S_2}
 \right],
\label{eq:cthetalens}
\ea
where $\Delta \frac{dN (S_1,S_2,\theta)}{d S_1 d S_2}$ is as given in
Eq.\ \ref{eq:dndssquare}.  Note that at this point we are ignoring
intrinsic correlations of the submillimeter objects themselves.  While
this is a concern at high $\ell$ values $\gtrsim 500$ corresponding to
the galaxy-galaxy correlation scale (Scott and White 1998), it
introduces little uncertainty into the analysis at $\ell$s
corresponding to cluster correlations.

If one imposes no upper limit on the flux of the point sources, then
the lensing cannot add structure to the angular correlations of point
sources.  This is due to the conservation of total flux behind the
lenses, Eq.\ \ref{eq:sfluxcon}, which insures that statistically, the
average flux from a lensed pixel will be equal to that of unlensed
areas of the sky.  Unlike Poisson noise, imposing a finite flux cut by
removing point sources actually {\it increases} this contribution to
anisotropies as it skews the mean flux from lensed pixels to lower
values.  This introduces a flux anisotropy where there previously was
none, adding power on the angular scale given by the clustering of the
lenses.

We can calculate the magnitude of this effect
from the Legendre transform of Eq.\ \ref{eq:cthetalens}.
\ba
C_\ell^I &=& w_\ell 
\left\{
\left[
\left(
\frac{1}{A_{\rm min}} -
\frac{1}{A_{\rm max}} 
\right)^2
- 
\left(
\frac{1}{2 A_{\rm min}^2} -
\frac{1}{2 A_{\rm max}^2} 
\right)^2
\right] 
S_{\rm cut}^2  \left. \frac{d N_L}{d S_1} \right|_{S = S_{\rm cut}}
\right. 
\int_0^{S_{\rm cut}} d S_2 S_2
\frac{d N_L}{d S_2}
\nonumber \\
&-&
\left(\frac{1}{A_{\rm min}^2} - 
      \frac{1}{A_{\rm max}^2} \right) 
\int_0^{S_{\rm cut}} d S_1 S_1
\frac{d N_L}{d S_1} 
\int_0^{S_{\rm cut}} d S_2 S_2
\frac{d N_L}{d S_2} 
\nonumber \\
&+& 
\left.
\left( 
\int_0^{S_{\rm cut}} d S_1 S_1
 \int_{A_{\rm min}}^{A_{\rm max}} \frac{dA_1}{A_1^4}  
	\frac{dN_L}{dS_1} (S_1/A)
\right)
\left(
\int_0^{S_{\rm cut}} d S_2 S_2
  \int_{A_{\rm min}}^{A_{\rm max}} \frac{dA_2}{A_2^4}  
	\frac{dN_L}{dS_2} (S_2/A)
\right)
\right\},
\ea
where $w_\ell \approx 2 \pi \theta_0^{-\gamma-1} (\ell^{-\gamma-3})$.
This expression is only at $\ell$ values below
the angular size of a cluster at a typical redshift $\sim 0.5$, 
and at $\ell$ values above which a power law fit to the angular 
correlation function of galaxies is appropriate 
($5^\circ \gtrsim \theta
\gtrsim 1'$  or $10 \lesssim \ell \lesssim 1000$)
(Tadros, Efstathiou, \& Dalton 1998).

\begin{figure}
\centerline{ 
\psfig{file=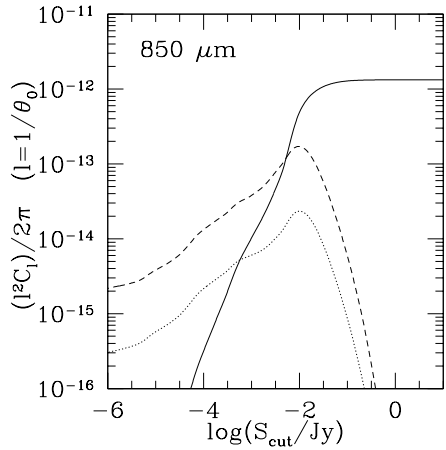,width=2.4in}}
\caption{
Contribution of lensing-induced structure to
measurements of
$\ell^2 C_\ell/2 \pi$ as a function of flux cut at $\ell = 1/\theta_0$,
the correlation angle. 
The solid line is the
Poisson noise of the unlensed distribution assuming $\theta_0 = 
1.1^\circ$, the dashed line
is the anisotropy due to lensing in the slowly evolving cluster model,
and the dotted line is the anisotropy in the FBC evolution model.
}
\label{fig:comparem}
\end{figure}

In Fig.\ \ref{fig:comparem} we plot this contribution to the
anisotropy of unresolved sources along with the Poisson noise
anisotropy at $\ell = 1/(1.1^\circ) = 27$.  As $C_\ell$
is proportional to $(\frac{dN_L}{dS})^2$, the slowly evolving 
lensing model gives a contribution $\sim 25$ greater than the
$\alpha = 2.9$ model.  Note also that this comparison
is only valid at this particular $\ell$ value as the Poisson noise
and lensing anisotropy have different scalings as a function of $\ell$,
with $\ell^2 C_\ell$ of the lensing contribution going as $\ell^{-\gamma-1}$,
and $\ell^2 C_\ell$ of Poisson noise going as $\ell^2$.  Thus,
at higher $\ell$ values, Poisson noise will be more dominant and will
be even more enhanced at very high $\ell$ values by galaxy-galaxy 
correlations, while at lower values the lensing anisotropy will play a larger
role.  For reference, the expected microwave background contribution
is given by $\ell^2 C_\ell/2 \pi \approx 10^{-10}$ for $\ell \lesssim 1000$.

\begin{figure}
\centerline{ 
\psfig{file=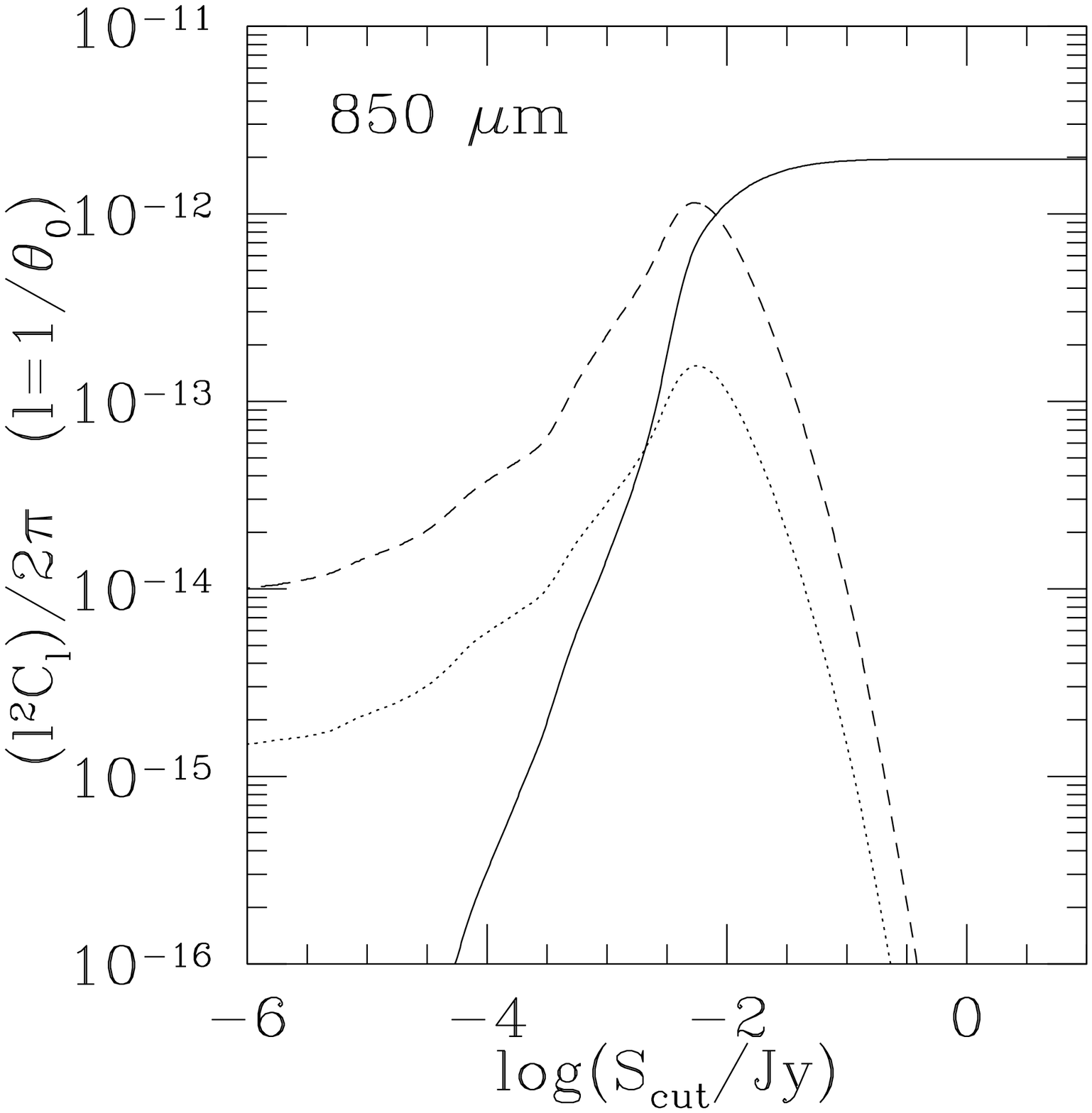,width=2.4in}}
\caption{
Contribution of lensing-induced structure
to measurements of
$\ell^2 C_\ell/ 2 \pi$  at $\ell = 1/\theta_0$.
as a function of flux cut, for a model 
($L{\rm cut} =  3 \times 10^{10} \;h^{-2} L_\odot$)
with a steeper source count slope near the flux cut 
The curves are as in Fig.\ \protect\ref{fig:comparem}.
}
\label{fig:compareh}
\end{figure}

In Fig.\ \ref{fig:compareh} we plot the same quantities, but now for
the $L_{\rm cut} = 3 \times 10^{10} \;h^{-2} L_\odot$ submillimeter
model.  While this model slightly overproduces the background flux and
most likely overpredicts both lensing and Poisson anisotropies, it
nevertheless serves to illustrate the sensitivity of our results to
the slope of the number counts near the flux-cut threshold.  While
Poisson noise is boosted only by a factor of two in this model, the
lensing anisotropy is increased by more than an order of magnitude by
a change in this slope.  Nevertheless, even in this extreme model, the
FBC lensing anisotropy falls short of the Poisson noise.

\begin{figure}
\centerline{ 
\psfig{file=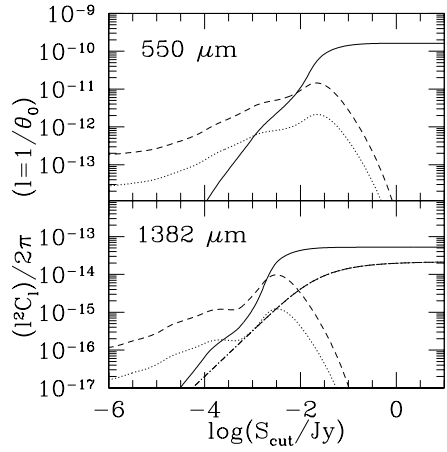,width=2.4in}}
\caption{
Contribution of lensing-induced structure
to measurements of
$\ell^2 C_\ell/ 2 \pi$ at $\ell = 1/\theta_0$
as a function of flux cut at two different
wavelengths.  The upper panel corresponds to the $550 \; \mu$m, 545 GHz
Planck channel, with the curves as in Fig.\ \protect\ref{fig:comparem}.
The lower panel is the same comparison for the $1382 \; \mu$m, 216 GHz
Planck channel.  The solid, short-dashed, and dotted curves are as before,
while the additional long-dashed curve is the simple extrapolation
by Scott and White (1998).
}
\label{fig:freq}
\end{figure}

Finally, in Fig.\ \ref{fig:freq} we compare Poisson noise and noise
induced by lensing at $550 \; \mu$m and $1382 \; \mu$m, which
correspond to the Planck high-frequency channels above and below the
$850 \mu$m channel we have considered so far.  Again our results
indicate that Poisson noise is dominant over structure induced by
lensing for realistic models of clustering.  Note, however, that the
level of submillimeter anisotropies relative to microwave background
anisotropies are drastically effected by changes in wavelength.  At
545 GHz we expect background measurements to be dominated by point
source contributions, that is, the far-infrared background.  At 217
GHz, however, point source contributions are much reduced, becoming
comparable to microwave anisotropies only at scales smaller than $\ell
\sim 20/\theta_0 = 1000.$ As a side note we point out that our
semi-empirical model shows a much sharper decrease in Poisson noise in
the 1 to 10 mJy range at 217 GHz than the simple extrapolation by
Scott \& White (1998), due to the sharper turnover of $\frac{dN}{dS}$
in the TSB model.  Thus high-$\ell$ microwave anisotropy measurements
at this frequency may gain more from a careful subtraction of point
sources than previously expected.

\section{Conclusions}

While the surface density of submillimeter objects was previously
unknown to within a factor of a thousand, recent observations
have given us our first glimpse into the submillimeter universe.  
This breakthrough has allowed authors to quantify the
level of background emission, confusion limits on future
observations, and the effects of lensing on the number counts
of resolved objects.

In this work, we have examined the effect of lensing by clusters on
the angular distribution of submillimeter sources.  While lensing does
not affect the total flux, it can cause a flux decrement of unresolved
lensed sources in flux-limited samples.  In the case of lensing by
clusters, this flux decrement is correlated on scales corresponding to
the cluster correlation length $\sim 1.1^\circ$, and thus represents a
possible contribution to background anisotropies at microwave and
far-infrared wavelengths.

Using a simple model for lensing by clusters and a semi-empirical
model for the evolution of lensed objects, we have quantified this
contribution, relating it to the anisotropy due to the discreteness of
the unresolved source distribution.  While slowly evolving 
cluster models predict structure comparable to Poisson noise on
angles $\sim 1.1^\circ$, more likely cluster models predict
anisotropies to be dominated by Poisson noise.  The degree of
structure induced in an $\Omega_{\rm M} < 1$ universe is expected to
be more than for the ``best guess'', $\Omega_{\rm M} = 1$ model, but
still fall short of the slowly evolving case we have considered.

Throughout this work we have assumed a uniform distribution of
submillimeter objects and have not addressed the question of the
clustering of the sources themselves. This clustering length is much
shorter than that of galaxy clusters and the clustering weaker, but
it nevertheless may represent a significant contribution at small
angles.  At larger angles this effect is negligible, however, and thus
the issue of structure in the far-infrared background itself does not
effect the conclusions drawn here.  It is with some confidence then
that we can say that experiments measuring microwave background
anisotropies and far-infrared background anisotropies due to the
clustering of unresolved sources will be able to remove all resolved
point sources without inducing structure in unresolved sources from
lensing by galaxy clusters.

\section*{Acknowledgements}
We appreciate valuable input from Rychard Bouwens, Tom Broadhurst, 
R. Benton Metcalf, D. M. Sherfesee, and Edward L. Wright.

\fontsize{10}{12pt}\selectfont

\newpage

\end{document}